\documentclass{appolb}
\usepackage[dvips]{graphicx}
\usepackage{amsmath}
\usepackage{amsfonts}
\usepackage{amssymb}
\usepackage{longtable} 
\usepackage{color}

\allowdisplaybreaks[4] 

\newcommand{\Chr}[3]{\mbox{\small\( \begin{Bmatrix}#1\\#2#3\end{Bmatrix}\)}}

\newcommand{\df}{\ {\overset {\rm def} =}\ }

\newcommand{\dril}[2]{{{\rm d} {#1}} / {{\rm d} {#2}}}

\newcommand{\lie}[2] {{\underset {#1} {\pounds}} {#2}}

\eqsec  

\begin{document}

\title{Spacetimes with no position drift}

\author{Andrzej Krasi\'nski
\address{N. Copernicus Astronomical Centre, Polish Academy of Sciences, \\
Bartycka 18, 00 716 Warszawa, Poland} \\
email: akr@camk.edu.pl}

\maketitle
\begin{abstract}
This paper compares three criteria for a spacetime to be free of position drift:
those by Hasse and Perlick (HP), Krasi\'nski and Bolejko (KB) and Korzy\'nski
and Kopi\'nski (KK). A spacetime having no position drift means that every
observer sees all light sources in unchanging directions. The following is
shown: (1) The HP criterion is a necessary condition for the KK criterion to
apply. (2) If the spacetime metric obeys the Einstein equations with a perfect
fluid source, then another necessary condition for the KK criterion is the Weyl
tensor being zero. (3) Result (2) points to the Stephani metric, so it is shown
that this metric obeys an equation which is still one more necessary condition
for the KK criterion. (4) The general Szekeres metrics become drift-free by the
KK criterion only in the Friedmann limit. (5) The HP and KB criteria coincide,
and the HP zero-drift condition imposes on the Stephani metric the same
restriction as found by Krasi\'nski and Bolejko (KB). The relations between the
three criteria are displayed and compared in a diagram.

\end{abstract}

\section{Motivation and summary}\label{intro}

Some time ago, this author noted that in an inhomogeneous Universe a generic
observer should see distant light sources drift across the sky, unlike in
Robertson -- Walker metrics. The drift is caused by the shearing and rotating
motion of the cosmic matter that sweeps the light rays passing through it, so
the direction from which a ray reaches an observer changes (relative to other
light sources) with the observer's time. The no-drift condition was that light
rays coming at different times to an observer ${\cal O}$ from the emitter ${\cal
E}$ intersect on their way always the same intermediate world lines of cosmic
matter. Equations governing the drift were derived for the class I Szekeres
models \cite{Szek1975} -- \cite{PlKr2006} in Ref. \cite{KrBo2011} and for the
Barnes \cite{Barn1973} and the expanding Stephani models \cite{Step1967} in
Refs. \cite{Kras2011} and \cite{Kras2012}. The angular rate of the drift
calculated in an exemplary Lema\^{\i}tre \cite{Lema1933} -- Tolman
\cite{Tolm1934} (L--T) model ($\approx 10^{-6}$ arc seconds per year
\cite{KrBo2011}) was on the verge of detectability for the
then-being-constructed Gaia satellite. A no-drift condition defined in a way
equivalent to that of Ref. \cite{KrBo2011} was discussed earlier by Hasse and
Perlick \cite{HaPe1988} without reference to explicit solutions of Einstein's
equations.

Recently, Korzy\'nski and collaborators published a series of papers in which
they discussed a (seemingly) differently defined drift in a general spacetime
\cite{KoKo2018} -- \cite{KMSe2021}. They derived a formula for the position
drift, not just a zero-drift condition. Their formula is the Fermi -- Walker
derivative along the observer world line of the unit direction vector to the
light source. The aim of the present paper is to relate to each other the sets
of results by the Korzy\'nski -- Kopi\'nski (KK) \cite{KoKo2018}, Hasse --
Perlick (HP) \cite{HaPe1988} and Krasi\'nski -- Bolejko (KB)
\cite{KrBo2011,Kras2011} teams.

In Sec. \ref{SNT}, the semi-null tetrad defined at a point by an observer
velocity $u^{\alpha}$ and a past-directed null vector $p^{\alpha}$
\cite{KoKo2018} is described. Formulae are given for the tetrad
${e_i}^{\alpha}$, the inverse tetrad ${e^i}_{\alpha}$, the scalar metric
$\eta_{i j}$, the inverse scalar metric $\eta^{i j}$, and for the tetrad
components of $u^{\alpha}$ and $p^{\alpha}$.

In Sec. \ref{posdrift}, the position drift formulae are quoted from
\cite{KoKo2018} and explained.

In Sec. \ref{implizerdri} it is noted that zero drift by the HP definition is a
necessary condition for zero drift by the KK definition. Then, in the same
section and Appendix A it is shown that if the cosmic matter is a perfect fluid
and the metric obeys the Einstein equations with this fluid as a source, then
zero drift in the KK sense implies zero Weyl tensor.

All conformally flat perfect fluid metrics are known, they are the Stephani
metrics \cite{Step1967,Step2003}. It is shown in Sec. \ref{conflat} and Appendix
B that they obey Eqs. (\ref{5.5}) and (\ref{4.4}), which are other necessary
conditions for the KK zero drift.

In Sec. \ref{Szek} it is shown that the  Szekeres metrics are drift-free in the
KK sense only in the Friedmann limit. This agrees with the result of Ref.
\cite{KrBo2011}.

In Sec. \ref{Perlick} it is shown that the KB definition of zero drift
\cite{KrBo2011,Kras2011} coincides with HP's \cite{HaPe1988}. It is also
verified that the HP definition applied to the expanding Stephani metric leads
to the same (axially symmetric) subcase as the KB definition.

In Sec. \ref{compa} the relations between the HP, KB and KK approaches and their
applications to the Stephani metric are explained and discussed in more detail,
and displayed in a diagram.

Section \ref{summa} contains a brief summary of all the results.

The signature $(+ - - -)$ will be used throughout most of the paper, except
where comparisons with Ref. \cite{KoKo2018} are made.

\section{The semi-null tetrad}\label{SNT}

\setcounter{equation}{0}

The spacetimes considered here contain world lines of light emitters ${\cal E}$
and observers ${\cal O}$ with four-velocities $u_{\cal E}^{\alpha}$ and $u_{\cal
O}^{\alpha}$, and light rays with tangent vectors $p^{\alpha}$. Occasionally, we
refer to a third observer with four-velocity $U^{\alpha}$.

At each point where an observer world line and a light ray intersect, a {\it
semi-null tetrad} (SNT) \cite{GKSe2019} of vectors can be introduced, on which
tensors can be projected. Tetrad indices will be denoted by small latin letters
$i, j, \dots = 0, 1, 2, 3$, tensor indices by Greek letters. Numerical values of
tetrad indices will have a hat above them. The tetrad indices running through
the two values $\widehat {1}$ and $\widehat {2}$ will be denoted by $A, B, C,
\dots$.

The contravariant vectors of the SNT are chosen as follows:
\begin{eqnarray}
{e_{\widehat {0}}}^{\alpha} &=& u^{\alpha}, \label{2.1}\\
{e_A}^{\alpha}, && A = 1, 2,\label{2.2} \\
{e_{\widehat {3}}}^{\alpha} &=& p^{\alpha}, \label{2.3}
\end{eqnarray}
where $u^{\alpha}$ is a timelike unit vector ($u^{\alpha} u_{\alpha} = 1$),
$p^{\alpha}$ is a null vector ($p^{\alpha} p_{\alpha} = 0$), and
${e_A}^{\alpha}$ are two spacelike unit vectors, orthogonal to each other and to
both  $u^{\alpha}$ and $p^{\alpha}$, so $g_{\alpha \beta} {e_A}^{\alpha}
{e_B}^{\beta} = - \delta_{AB}$, ${e_A}^{\alpha} p_{\alpha} = {e_A}^{\alpha}
u_{\alpha} = 0$, not otherwise specified. The ${e_A}^{\alpha}$ are defined at
${\cal O}$ and parallely transported along $p^{\alpha}$.

The tetrad metric $\eta_{i j} = g_{\alpha \beta} {e_i}^{\alpha} {e_j}^{\beta}$
is then
\begin{eqnarray}
&& \eta_{\widehat {0} \widehat {0}} = 1, \quad \eta_{\widehat {0} \widehat {1}}
= \eta_{\widehat {0} \widehat {2}} = 0, \quad \eta_{\widehat {0} \widehat {3}} =
u_{\rho} p^{\rho},  \label{2.4}\\
&& \eta_{\widehat {1} \widehat {1}} = \eta_{\widehat {2} \widehat {2}} = -1,
\label{2.5} \\
&& \eta_{\widehat {1} \widehat {2}} = \eta_{\widehat {1} \widehat {3}} =
\eta_{\widehat {2} \widehat {3}} = \eta_{\widehat {3} \widehat {3}} = 0.
\label{2.6}
\end{eqnarray}
The inverse metric to $\eta_{i j}$ is
\begin{eqnarray}
&& \eta^{\widehat {0} \widehat {0}} = \eta^{\widehat {0} \widehat {1}} =
\eta^{\widehat {0} \widehat {2}} = 0, \quad \eta^{\widehat {0} \widehat {3}} =
1 / u_{\rho} p^{\rho},  \label{2.7}\\
&& \eta^{\widehat {1} \widehat {1}} = \eta^{\widehat {2} \widehat {2}} = -1,
\label{2.8} \\
&& \eta^{\widehat {1} \widehat {2}} = \eta^{\widehat {1} \widehat {3}} =
\eta^{\widehat {2} \widehat {3}} = 0, \label{2.9} \\
&&\eta^{\widehat {3} \widehat {3}} = -1 / \left(u_{\rho} p^{\rho}\right)^2.
\label{2.10}
\end{eqnarray}
Thus, the covariant tetrad ${e^i}_{\alpha} = \eta^{i s} g_{\alpha \rho}
{e^s}_{\rho}$ is
\begin{eqnarray}
{e^{\widehat {0}}}_{\alpha} &=& p_{\alpha} / \left(u_{\rho}
p^{\rho}\right), \label{2.11} \\
{e^{\widehat {1}}}_{\alpha} &=& -e_{{\widehat {1}} \alpha}, \quad {e^{\widehat
{2}}}_{\alpha} = -e_{{\widehat {2}} \alpha}, \label{2.12} \\
{e^{\widehat {3}}}_{\alpha} &=& u_{\alpha} / \left(u_{\rho} p^{\rho}\right) -
p_{\alpha} / \left(u_{\rho} p^{\rho}\right)^2. \label{2.13}
\end{eqnarray}
The tetrad components of $u^{\alpha}$ and $p^{\alpha}$ are
\begin{eqnarray}
u_{\widehat {0}} &=& 1, \quad u_{\widehat {3}} = u_{\rho} p^{\rho}, \quad
u_{\widehat {1}} = u_{\widehat {2}} = 0, \label{2.14} \\
u^{\widehat {0}} &=& 1, \quad u^{\widehat {1}} = u^{\widehat {2}} = u^{\widehat
{3}} = 0, \label{2.15} \\
p_{\widehat {0}} &=& u_{\rho} p^{\rho}, \quad p_{\widehat {1}} = p_{\widehat
{2}} = p_{\widehat {3}} = 0, \label{2.16} \\
p^{\widehat {0}} &=& p^{\widehat {1}} = p^{\widehat {2}} = 0, \quad p^{\widehat
{3}} = 1. \label{2.17}
\end{eqnarray}

\section{Position drift according to Ref. [12]}\label{posdrift}

\setcounter{equation}{0}

We recall the definitions of the basic notions introduced in Ref.
\cite{KoKo2018}.

At a point ${\it P}$ of a manifold a future-directed timelike vector
$u^{\alpha}$ (which is the four-velocity of an observer) and a past-directed
null vector $p^{\alpha}$ (which is tangent to a light ray reaching the observer)
define the unit spacelike vector $r^{\alpha}$ pointing from ${\it P}$ to the
light source:
\begin{equation}\label{3.1}
r^{\alpha} = u^{\alpha} - p^{\alpha}/\left(p^{\rho}u_{\rho}\right).
\end{equation}

The {\it Fermi -- Walker (FW) transport} of a vector $V^{\alpha}$ along a unit
timelike vector field $U^{\alpha}$ ($U^{\alpha} U_{\alpha} = 1$) is such, by
which the component of $V^{\alpha}$ lying in the $\{U^{\alpha},
\dot{U}^{\alpha}\}$ plane, where $\dot{U}^{\alpha} \df U^{\rho}
{U^{\alpha}};_{\rho}$, remains in this plane (so $V^{\alpha}$ does not rotate
around $U^{\alpha}$). Then $V^{\alpha}$ obeys \cite{Rind2011}
\begin{equation}\label{3.2}
U^{\rho} V^{\alpha};_{\rho} = \left(V^{\rho} U_{\rho}\right) \dot{U}^{\alpha} -
\left(V^{\rho} \dot{U}_{\rho}\right) U^{\alpha} \df {\overset * V}{}^{\alpha}.
\end{equation}
If $U^{\alpha}$ is geodesic, then $\dot{U}^{\alpha} =  {\overset * V}{}^{\alpha}
= 0$ and the FW and parallel transports coincide. The {\it Fermi-Walker
derivative} $\delta_U$ is defined by
\begin{equation}\label{3.3}
(\delta_U V)^{\alpha} \df U^{\rho} V^{\alpha};_{\rho} - {\overset *
V}{}^{\alpha},
\end{equation}
so that $(\delta_U V)^{\alpha} = 0$ when $V^{\alpha}$ is FW-transported along
$U^{\alpha}$. Note that $(\delta_U U)^{\alpha} = 0$.

The FW transport and FW derivative can be generalised to arbitrary tensor fields
\cite{KoKo2018} and then it follows that $\delta_U g_{\alpha \beta} = 0$ for any
metric $g_{\alpha \beta}$ and any vector field $U^{\alpha}$. From this it
follows that $\delta_U \left(g_{\alpha \beta} P^{\alpha} Q^{\beta}\right) = 0$
for any FW-transported vector fields $P^{\alpha}$ and $Q^{\beta}$, so the FW
transport preserves the angle between $P^{\alpha}$ and $Q^{\beta}$.

The FW derivative along the observer's world line will be denoted $\delta_{\cal
O}$.

Let a geodesic $G$ belong to a one-parameter family $F_G$ of geodesics, and let
$p^{\alpha}$ be the vector field tangent to $G$. Then the field of vectors
$\xi^{\alpha}$ pointing from points on $G$ toward neighbouring geodesics in
$F_G$ is called {\it geodesic deviation} vector field and obeys the {\it
geodesic deviation equation}
\begin{equation}\label{3.4}
{\cal G} \left[\xi\right]^{\mu} \df \nabla_p \left(\nabla_p \xi^{\mu}\right) -
{R^{\mu}}_{\alpha \beta \nu} p^{\alpha} p^{\beta} \xi^{\nu} = 0,
\end{equation}
where $\nabla_p \xi^{\mu} \df p^{\rho} {\xi^{\mu}};_{\rho}$ and
${R^{\mu}}_{\alpha \beta \nu}$ is the curvature tensor. The definition of
$\xi^{\mu}$ implies
\begin{equation}\label{3.5}
\left(\nabla_{\xi} p\right)^{\mu} = \left(\nabla_p \xi\right)^{\mu},
\end{equation}
which means that the fields $p^{\alpha}$ and $\xi^{\alpha}$ are surface-forming.

The definition of geodesic deviation applies to any bundle of geodesics, but in
the following the geodesics will be null.

Imagine a bundle of past-directed rays emanating from the same observation event
and let $\lambda$ be the affine parameter on them. Take one ray as the reference
and consider the deviation vectors $\xi^{\alpha}$ along it. What matters in
calculating the position drift of an observed source are the projections $\xi^A
= {{e^A}}_{\mu} \xi^{\mu}$ of $\xi^{\mu}$ on the 2-dimensional planes orthogonal
to $p^{\mu}$ and to $u_{\cal O}^{\mu}$. Since the rays converge to the same
point, we have $\xi^A(\lambda_{\cal O}) = 0$ at the observer, and then
$\xi^A(\lambda)$ at any other $\lambda$ is uniquely defined by ${e^A}_{\mu}
\left(\nabla_{\xi} p\right)^{\mu} = {e^A}_{\mu} \left(\nabla_p \xi\right)^{\mu}
= \dril {\xi^A} {\lambda}$ (we assume there are no caustics between the observer
and the light source). Since (\ref{3.4}) is linear in $\xi^{\mu}$, the {\it
Jacobi matrix} ${{\cal D}^A}_B$ exists such that
\begin{equation}\label{3.6}
\xi^A(\lambda) = {{\cal D}^A}_B(\lambda) \nabla_p \xi^B(\lambda_{\cal O}).
\end{equation}
The ${{\cal D}^A}_B$ maps vectors tangent to the past light cone at ${\cal O}$
to vectors attached at other points along the rays. Since the mapping is 1 -- 1,
the inverse mapping also exists.

\begin{figure}[h]
\begin{center}
\includegraphics[scale = 0.6]{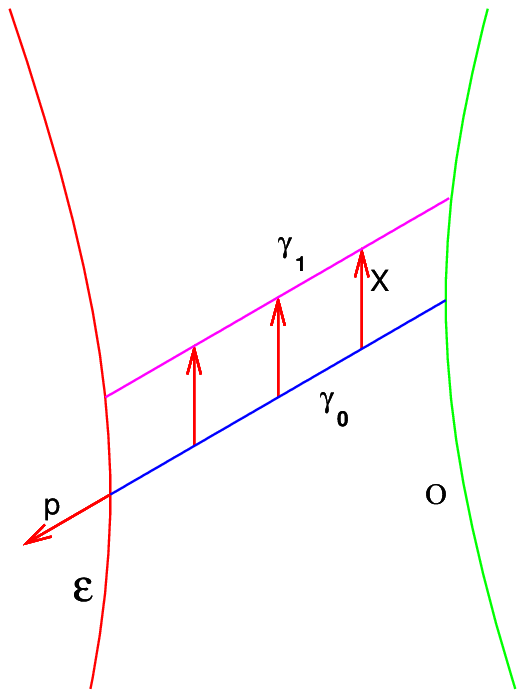}
\caption{The emitter ${\cal E}$ keeps sending light rays to the observer ${\cal
O}$; two such rays, $\gamma_0$ and $\gamma_1$, are shown. $X$ is the geodesic
deviation vector field along $\gamma_0$ and $p$ is the past-directed tangent
vector field to $\gamma_0$. (This is simplified Fig. 8 from Ref.
\cite{KoKo2018}. )}
 \label{obstimevec}
\end{center}
\end{figure}

Now let a source ${\cal E}$ send rays $\gamma_0, \gamma_1, \dots$ that intersect
the observer world line ${\cal O}$, as in Fig. \ref{obstimevec}. The
corresponding geodesic deviation field $X^{\mu}$, called {\it observation time
vector} \cite{KoKo2018}, is collinear with the observer velocity $u_{\cal
O}^{\alpha}$ at ${\cal O}$ and with the emitter velocity $u_{\cal E}^{\alpha}$
at ${\cal E}$. We choose the affine parameters on the rays so that $\lambda =
\lambda_{\cal O}$ at all points of intersection with the observer world line,
and $\lambda = \lambda_{\cal E}$ at all points of intersection with the emitter
world line. Then $X^{\mu}$ obeys ${\cal G} [X]^{\mu} = 0$ and the initial
conditions \cite{KoKo2018}
\begin{equation}\label{3.7}
X^{\mu}(\lambda_{\cal O}) = u^{\mu}_{\cal O}, \qquad X^{\mu}(\lambda_{\cal E}) =
\frac {u^{\mu}_{\cal E}} {1 + z},
\end{equation}
where $z$ is the redshift along $\gamma_0$ between ${\cal E}$ and ${\cal O}$. We
split $X^{\mu}$ as follows
\begin{equation}\label{3.8}
X^{\mu} = \widehat{u}^{\mu}_{\cal O} + m^{\mu} + \phi^{\mu},
\end{equation}
where $\widehat{u}^{\mu}_{\cal O}$ is the observer four-velocity parallely
transported along $p^{\mu}$ from ${\cal O}$ to the running point on $\gamma$,
while $m^{\mu}$ and $\phi^{\mu}$ obey
\begin{eqnarray}
{\cal G}[m]^{\mu} &=& {R^{\mu}}_{\alpha \beta \nu} p^{\alpha} p^{\beta}
\widehat{u}^{\nu}_{\cal O}, \label{3.9} \\
{\cal G}[\phi]^{\mu} &=& 0, \label{3.10}
\end{eqnarray}
with the initial conditions
\begin{eqnarray}
m^{\mu}(\lambda_{\cal O}) &=& 0, \label{3.11} \\
\nabla_p m^{\mu}(\lambda_{\cal O}) &=& 0, \label{3.12} \\
\phi^{\mu}(\lambda_{\cal O}) &=& 0, \label{3.13} \\
\phi^{\mu}(\lambda_{\cal E}) &=& \frac {u^{\mu}_{\cal E}} {1 + z} -
\left.\widehat{u}^{\mu}_{\cal O}\right|_{\cal E} - m^{\mu}_{\cal E}.
\label{3.14}
\end{eqnarray}
Now $\phi^{\mu}(\lambda)$ can be calculated along $\gamma$ using (\ref{3.6})
\cite{KoKo2018}. Knowing all this, the position drift of the light source is
calculated to be \cite{KoKo2018}
\begin{eqnarray}\label{3.15}
\delta_{\cal O} r^A &=& \frac 1 {p_{\sigma} u_{\cal O}^{\sigma}}\ {\cal
D}^{-1}{(\lambda_{\cal E})^A}_B \left(\frac {u^B} {1 + z} - \widehat{u}_{\cal
O}^B - m^B\right)_{\cal E} + w_{\cal O}^A, \nonumber \\
&\equiv& \frac 1 {p_{\sigma} u_{\cal O}^{\sigma}}\ {\cal D}^{-1}{(\lambda_{\cal
E})^A}_B \phi^B\left(\lambda_{\cal E}\right) + w_{\cal O}^A
\end{eqnarray}
where $w_{\cal O}^{\mu}$ is the observer's acceleration. The vector field
$U^{\alpha}$ defining the tetrad for (\ref{3.15}) need not coincide with
$u_{\cal E}^{\alpha}$, but if it does then $u_{\cal E}^B = 0$.

Since $\phi^{\mu}$ obeys (\ref{3.10}) and (\ref{3.13}), we can apply (\ref{3.6})
to it, then
\begin{equation}\label{3.16}
\delta_{\cal O} r^A = \frac 1 {p_{\sigma} u_{\cal O}^{\sigma}}\ \left(\nabla_p
\phi^A\right)_{\cal O} + w_{\cal O}^A.
\end{equation}

\section{Implications of $\delta_{\cal O} r^A = 0$ in
(3.16)}\label{implizerdri}

\setcounter{equation}{0}

The FW derivative preserves the scalar products of vectors, so $0 = \delta_{\cal
O} \left(r_{\mu} r^{\mu}\right) = 2 r_{\mu} \delta_{\cal O} r^{\mu}$, i.e.,
$\delta_{\cal O} r^{\mu}$ is orthogonal to $r^{\mu}$. It is also orthogonal to
$u_{\cal O}^{\mu}$ because $\delta_{\cal O} u_{\cal O}^{\mu} = 0$ (see the
remark under (\ref{3.3})). Consequently, when $\delta_{\cal O} r^A = 0$, the
whole 4-dimensional $\delta_{\cal O} r^{\mu} = 0$. Then, from the paragraph
under (\ref{3.3}) it follows that angles between the directions to any pair of
light sources will remain constant in observer's time -- which is the HP
\cite{HaPe1988} criterion for zero position drift. This means that the
spacetimes that are drift-free in the KK sense are necessarily drift-free also
in the HP sense. We will discuss the latter in Sec. \ref{Perlick}. We first
investigate a consequence of $\delta_{\cal O} r^A = 0$ in (\ref{3.16}).

In a zero-drift spacetime $X^{\mu}$ must be everywhere tangent to a surface
formed by the world-lines of the cosmic medium. Consequently, $X^A = 0$ and the
change of $X^{\mu}$ along $p^{\mu}$ must also be tangent to this surface; i.e.,
\begin{equation}\label{4.1}
{\nabla_p X}^{\mu} = p^{\rho} {X^{\mu}};_{\rho} = f X^{\mu} + g p^{\mu},
\end{equation}
where $f$ and $g$ are functions on the spacetime. The field $X^{\mu}$ obeys
(\ref{3.4}):
\begin{equation}\label{4.2}
\left(\nabla_p \nabla_p X\right)^{\mu} - {R^{\mu}}_{\alpha \beta \nu} p^{\alpha}
p^{\beta} X^{\nu} = 0.
\end{equation}
{}From (\ref{4.1}) we have ${\nabla_p \nabla_p X}^{\mu} = \left({\nabla_p f} +
f^2\right) X^{\mu} + \left(f g + {\nabla_p g}\right) p^{\mu}$, so
\begin{equation}\label{4.3}
\left({\nabla_p f} + f^2\right) X^{\mu} + \left(f g + {\nabla_p g}\right)
p^{\mu} = {R^{\mu}}_{\alpha \beta \nu} p^{\alpha} p^{\beta} X^{\nu}.
\end{equation}
Projecting this on $p^{\mu}$ we obtain ${\nabla_p f} + f^2 = 0$. Finally then
\begin{equation}\label{4.4}
\left(f g + {\nabla_p g}\right) p^{\mu} = {R^{\mu}}_{\alpha \beta \nu}
p^{\alpha} p^{\beta} X^{\nu}.
\end{equation}
We project (\ref{4.4}) on the ${e^A}_{\mu}$ vectors that obey ${e^A}_{\mu}
p^{\mu} = 0$. The result is
\begin{equation}\label{4.5}
{R^A}_{\alpha \beta \nu} p^{\alpha} p^{\beta} X^{\nu} = 0.
\end{equation}
But (\ref{4.5}) is not equivalent to (\ref{4.2}) -- it is only a subset of
consequences of (\ref{4.2}). Consequently, conclusions drawn from (\ref{4.5})
will not fully represent (\ref{4.2}), they will only be necessary conditions for
(\ref{4.2}). We will come back to this in Sec. \ref{conflat}.

We replace the coordinate summation indices with the tetrad summation indices
and recall that $p^{\alpha} = {e_{\widehat {3}}}^{\alpha}$. Then (\ref{4.5})
becomes
\begin{equation}\label{4.6}
{R^A}_{\widehat{3} \widehat{3} i}\ X^i = 0.
\end{equation}
The component $X^{\widehat 3}$ gives zero contribution to (\ref{4.6}) (and is
zero anyway), $X^A = 0$, so what remains of (\ref{4.6}) is ${R^A}_{\widehat{3}
\widehat{3} \widehat{0}} X^{\widehat{0}} = 0$. But from (\ref{2.15})
$X^{\widehat{0}} = u^{\widehat{0}} / (1 + z) \neq 0$. Thus, (\ref{4.6}) implies
${R^A}_{\widehat{3} \widehat{3} \widehat{0}} = 0$, and so, using (\ref{2.4}) --
(\ref{2.6}):
\begin{equation}\label{4.7}
{R^{A \widehat{0}}}_{\widehat{3} \widehat{0}} \equiv {R^A}_{\widehat{3}
\widehat{3} \widehat{0}}\ / u_{\rho} p^{\rho} = 0.
\end{equation}
For the Weyl tensor ${C^{\alpha \beta}}_{\gamma \delta}$ we have in tetrad
components \cite{PlKr2006}
\begin{equation}\label{4.8}
{R^{i j}}_{k l} = {C^{i j}}_{k l} - \frac 1 2 \delta^{i j r}_{k l s}
\left({{R}^s}_r - \frac 1 4\ {\delta^s}_r R\right) + \frac 1 {12}\ \delta^{i
j}_{k l} R,
\end{equation}
where ${R^i}_j$ is the Ricci tensor and $R = {R^s}_s$. Using (\ref{4.8}) in
(\ref{4.7}) we find:
\begin{equation}\label{4.9}
{R^{A \widehat{0}}}_{\widehat{3} \widehat{0}} = {C^{A \widehat{0}}}_{\widehat{3}
\widehat{0}} + \frac 1 2\ {R^A}_{\widehat {3}} = 0.
\end{equation}
This simplifies when the cosmic matter is a perfect fluid (this assumption
underlies all cosmological models) and the Einstein equations are obeyed
\begin{equation}\label{4.10}
{R^i}_j = \kappa \left[(\epsilon + p)u^i u_j - \frac 1 2 (\epsilon - p)
{\delta^i}_j\right],
\end{equation}
where $\kappa = 8 \pi G / c^4$, $\epsilon$ is the energy density and $p$ is the
pressure. In the frame (\ref{2.1}) -- (\ref{2.3}) $u^A = 0$ and
${\delta^A}_{\widehat {3}} = 0$, so ${R^A}_{\widehat {3}} = 0$, and
\begin{equation}\label{4.11}
{C^A}_{\widehat{3} \widehat{3} \widehat{0}} = {C^A}_{\alpha \beta \gamma}\
p^{\alpha} p^{\beta} u^{\gamma} = 0.
\end{equation}
This must hold for an arbitrary null vector $p^{\alpha}$, but only for that one
$u^{\alpha}$ which is tangent to the cosmic matter flow line. It is shown in
Appendix A that the requirement of (\ref{4.11}) holding for all null fields
$p^{\alpha}$ leads to the whole Weyl tensor being zero. Thus the conclusion is

{\bf Conclusion 1}

For spacetimes in which the cosmic fluid is perfect and obeys the Einstein
equations (\ref{4.10}) the direction drift as defined by (\ref{3.16}) is zero
for all comoving observers only if the spacetime is conformally flat. $\square$

The conformally flat perfect fluid metrics are all explicitly known, they are
the Stephani metrics \cite{Step2003,Step1967}. In general, the perfect fluid in
them moves with acceleration, but in the limit of zero acceleration the
expanding Stephani metric reduces to the general Robertson -- Walker (RW)
metric, while the expansion-free one trivializes to the Einstein universe which
is a subcase of RW; see Sec. \ref{conflat} here. Hence:

{\bf Conclusion 2}

The only perfect fluid spacetimes in which the fluid moves geodesically and in
which the direction drift as defined by (\ref{3.16}) is zero for all comoving
observers are the Robertson -- Walker ones:
\begin{equation}\label{4.12}
{\rm d} s^2 = {\rm d} t^2 - R^2(t) \left[\frac {{\rm d} r^2} {1 - k r^2} + r^2
\left({\rm d} \vartheta^2 + \sin^2 \vartheta {\rm d} \varphi^2\right)\right].
\end{equation}

\section{The Stephani [6, 15] metrics obey (3.16) with
$\delta_{\cal O} r^A = 0$}\label{conflat}

\setcounter{equation}{0}

{}From (\ref{3.16}), $\delta_{\cal O} r^A = 0$ implies
\begin{equation}\label{5.1}
w_{\cal O}^A = - \frac 1 {p_{\sigma} u_{\cal O}^{\sigma}}\ \left(\nabla_p
\phi^A\right)_{\cal O}.
\end{equation}
{}From (\ref{3.8}) we have $\phi^A = X^A - \widehat{u}_{\cal O}^A - m^A$. But
$\widehat{u}_{\cal O}^{\alpha}$ is parallely transported along $p^{\alpha}$, so
$\left(\nabla_p \widehat{u}_{\cal O}\right)^A = 0$, and from (\ref{3.12})
$\left(\nabla_p m\right)^A \left(\lambda_{\cal O}\right) = 0$. Consequently,
\begin{equation}\label{5.2}
\left(\nabla_p \phi\right)_{\cal O}^A = \left(\nabla_p X\right)_{\cal O}^A.
\end{equation}
In a drift-free spacetime $X^{\alpha} = u^{\alpha} / (1 + z)$ on each ray. Thus
\begin{equation}\label{5.3}
\left(\nabla_p X\right)_{\cal O}^A = \left.\frac {\left(\nabla_p u\right)^A} {1
+ z}\right|_{\cal O} - \frac {u^A_{\cal O}} {(1 + z)^2}\ \nabla_p z.
\end{equation}
But $u^A_{\cal O} = \left({e^A}_{\mu} u^{\mu}\right)_{\cal O} = 0$ and $z = 0$
at the observer. Therefore
\begin{eqnarray}
\left(\nabla_p X\right)_{\cal O}^A &=& \left(\nabla_p u\right)^A_{\cal O},
\label{5.4} \\
w_{\cal O}^A &=& - \frac 1 {p_{\sigma} u_{\cal O}^{\sigma}}\ \left(\nabla_p
u\right)^A_{\cal O}. \label{5.5}
\end{eqnarray}
Note that $u^A_{\cal O} = 0$ does not imply $(\nabla_p u)^A_{\cal O} = 0$.

We will show here that the Stephani metrics obey (\ref{5.1}), so zero Weyl
tensor and a perfect fluid source constitute together a sufficient condition for
(\ref{5.1}) to hold. For a while we switch to the signature $(- + + +)$ that was
used in Ref. \cite{KoKo2018}. The expanding Stephani metric is
\begin{equation}\label{5.6}
{\rm d} s^2 = - \left(\frac {F V,_t} V\right)^2 {\rm d} t^2 + \frac 1 {V^2}\
\left({\rm d} x^2 + {\rm d} y^2  {\rm d} z^2\right),
\end{equation}
where
\begin{equation}\label{5.7}
V = \frac 1 {R(t)}\ \left\{1 + \frac 1 4\ k(t) \left[\left(x - x_0(t)\right)^2 +
\left(y - y_0(t)\right)^2 + \left(z - z_0(t)\right)^2\right]\right\},
\end{equation}
the functions $R(t)$, $k(t)$, $x_0(t)$, $y_0(t)$ and $z_0(t)$ being all
arbitrary. It obeys the Einstein equations with a perfect fluid source, the mass
density $\rho$ and pressure $p$ are
\begin{equation}\label{5.8}
\kappa c^2 \rho = 3 C^2(t), \qquad \kappa p = - 3 C^2(t) + 2 C C,_t V / V,_t,
\qquad \kappa \df 8 \pi G / c^4,
\end{equation}
where the arbitrary function $C(t)$ is related to the other ones by
\begin{equation}\label{5.9}
C^2 = k R^2 + 1 / F^2.
\end{equation}
The velocity $u^{\alpha}$ and acceleration $w^{\alpha} = {u^{\alpha}};_{\rho}
u^{\rho}$ fields in (\ref{5.6}) are
\begin{equation}\label{5.10}
u^{\alpha} = \frac V {F V,_t}\ {\delta^{\alpha}}_0,
\end{equation}
\begin{equation}\label{5.11}
w^0 = 0, \qquad w^I = \frac V {V,_t}\ \left(V V,_{t I} - V,_t V,_I\right),
\end{equation}
where $I = 1, 2, 3, (x^1, x^2, x^3) = (x, y, z)$. (The spatial coordinate index
$I$ is not to be confused with the tetrad index $A$.)

The $A$ components in (\ref{5.5}) are projections of $w_{\cal O}^{\alpha}$ and
$\left(\nabla_p u\right)^{\alpha}$ on the vectors ${e_A}^{\alpha}$ of the tetrad
discussed in Sec. \ref{SNT}. Let $q^{\alpha}$ be any of the ${e_A}^{\alpha}$.
Since it is orthogonal at ${\cal O}$ to the $u^{\alpha}$ of (\ref{5.10}), it
must have $q^0 = 0$. Then, orthogonality to $p^{\alpha}$ in the metric
(\ref{5.6}) means
\begin{equation}\label{5.12}
p^1 q^1 + p^2 q^2 + p^3 q^3 = 0.
\end{equation}
The projection of $\left(\nabla_p u\right)^{\alpha}$ on $q^{\alpha}$ (i.e., one
of the $(\nabla_p u)^A$) at ${\cal O}$ is
\begin{equation}\label{5.13}
(\nabla_p u)^q_{\cal O} = g_{\alpha \beta} q_{\cal O}^{\alpha} (\nabla_p
u)^{\beta}_{\cal O} = \nabla_p \left(g_{\alpha \beta} q^{\alpha}
u^{\beta}\right)_{\cal O} = \frac 1 {V^2}\ \sum_{I = 1}^3 q^I_{\cal O}
\left(\nabla_p u\right)^I_{\cal O}.
\end{equation}
We have
\begin{eqnarray}
&& \left(\nabla_p u\right)^I_{\cal O} = \left(p^{\rho}
{u^I};_{\rho}\right)_{\cal O} = \left(p^{\rho} {u^I},_{\rho}\right)_{\cal O} +
\left(p^{\rho} \Chr {I\ } {0 \rho}\ u^0\right)_{\cal O}, \label{5.14} \\
&& u^I = 0 \qquad {\rm everywhere\ in\ the\ coordinates\ of\ (\ref{5.6})},
\label{5.15} \\
&& \Chr {I}\ {0 0} = {\displaystyle F^2 \frac {V,_t} V\ \left(V V,_{t I} -
V,_t V,_I\right)}, \label{5.16} \\
&& \Chr {I}\ {0 J} = - {\displaystyle \frac {V,_t} V\ {\delta^I}_J}.
\label{5.17}
\end{eqnarray}
{}From here
\begin{equation}\label{5.18}
\left(\nabla_p u\right)^I_{\cal O} = p^0 F \left(V V,_{t I} - V,_t V,_I\right) -
p^I / F.
\end{equation}
So finally, using (\ref{5.12}):
\begin{equation}\label{5.19}
(\nabla_p u)^q_{\cal O} = \frac {p^0 F} {V^2}\ \sum_{I = 1}^3 q^I_{\cal O}
\left(V V,_{t I} - V,_t V,_I\right).
\end{equation}
In the signature $(- + + +)$ $u_0 = - F V,_t / V$ and $p^{\sigma} u_{\sigma} = -
p^0 F V,_t / V$, so
\begin{equation}\label{5.20}
\frac 1 {p_{\sigma} u^{\sigma}_{\cal O}} (\nabla_p u)^q_{\cal O} = - \frac 1 {V
V,_t}\ \sum_{I = 1}^3 q^I_{\cal O} \left(V V,_{t I} - V,_t V,_I\right).
\end{equation}
For the acceleration vector (\ref{5.11}) we have
\begin{equation}\label{5.21}
w^q = g_{\alpha \beta} q^{\alpha} w^{\beta} = \frac 1 {V V,_t}\ \sum_{I = 1}^3
q^I_{\cal O} \left(V V,_{t I} - V,_t V,_I\right),
\end{equation}
so (\ref{5.5}) is fulfilled. But we recall: the Stephani metric still has to
obey the remaining equations in the set (\ref{4.2}). So far, we have only made
use of (\ref{4.5}), which is a necessary condition for (\ref{4.2}), but is not
equivalent to (\ref{4.2}). Unfortunately, the explicit form of (\ref{4.2}) is
complicated and this author was not able to simplify it to a readable form.
Therefore, we will rely on the observation made in the first paragraph of Sec.
\ref{implizerdri}: the spacetimes that are drift-free in the KK sense are
necessarily drift-free also in the HP sense. What the HP criterion implies for
the expanding Stephani metric is shown in Sec. \ref{Perlick}.

We now repeat the same consideration with the expansion-free Stephani metric
\cite{Step2003}, still in the $(- + + +)$ signature and in the notation of Ref.
\cite{Kras1981}:\footnote{This expansion-free metric is not interesting for
cosmology, and this discussion is added only for completeness. It has no
consequences for the main topic of this paper.}
\begin{eqnarray}
{\rm d} s^2 &=& - D^2 {\rm d} t^2 + \frac {{\rm d} r^2} {1 + K r^2} + r^2
\left({\rm d} \vartheta^2 + \sin^2 \vartheta {\rm d} \varphi^2\right),
\label{5.22} \\
D &=& r V + E(t) \sqrt{1 + K r^2} +s, \label{5.23} \\
V &=& A(t) \sin \vartheta \cos \varphi + B(t) \sin \vartheta \sin \varphi + C(t)
\cos \vartheta, \label{5.24}
\end{eqnarray}
where $s = 1$ or 0, $K$ is an arbitrary constant, $A, B, C$ and $E$ are
arbitrary functions of $t$, and $(x^0, x^1, x^2, x^3) = (t, r, \vartheta,
\varphi)$. For this metric we have
\begin{eqnarray}
u^{\alpha} &=& \frac 1 D\ {\delta^{\alpha}}_0 , \label{5.25} \\
w^0 &=& 0, \qquad w^1 = g^{11} \frac {D,_r} D = \frac {1 + K r^2} D\ \left(V +
\frac {E r} {\sqrt{1 + K r^2}}\right), \label{5.26} \\
w^2 &=& \frac {V,_{\vartheta}} {D r}, \qquad w^3 = \frac {V,_{\varphi}} {D r
\sin^2 \vartheta}. \label{5.27}
\end{eqnarray}
As before, $q^0 = 0$ and $p^{\alpha} q_{\alpha} = 0$ implies an equation similar
to (\ref{5.12}), but it will not be useful this time. The projection of
$w^{\alpha}$ on $q^{\alpha}$ is
\begin{equation}\label{5.28}
w^q = g_{\alpha \beta} q^{\alpha} w^{\beta} = q^1 \frac {D,_r} D + q^2 \frac {r
V,_{\vartheta}} D + q^3 \frac {r V,_{\varphi}} D.
\end{equation}
Here we have $p_{\sigma} u^{\sigma} = p^0 u_0 = - D p^0$ and
\begin{eqnarray}
\left(\nabla_p u\right)^1_{\cal O} &=& g^{11} p^0 D,_r, \label{5.29} \\
\left(\nabla_p u\right)^2_{\cal O} &=& p^0 \frac {V,_{\vartheta}} r, \qquad
\left(\nabla_p u\right)^3_{\cal O} = p^0 \frac {V,_{\varphi}} {r \sin^2
\vartheta}, \label{5.30}
\end{eqnarray}
and with this $w^q = - \left(\nabla_p u\right)^q_{\cal O} / p_{\sigma}
u^{\sigma}$, which agrees with (\ref{5.5}). But this metric obeys the HP
criterion P5 for drift absence (see Sec. \ref{Perlick}) only in the special
cases $E = 0$, $K = 0$ and $V = 0$; the last one is the de Sitter metric.

For completeness, in Appendix B it is shown that the expanding Stephani metric
obeys Eq. (\ref{4.4}) with $f = 0$.

\section{Conditions for drift absence in the Szekeres metrics}\label{Szek}

\setcounter{equation}{0}

Let us now verify whether (\ref{5.5}) can hold in any of the Szekeres metrics.
Since $w^{\alpha} = 0$ in them, the cosmic velocity field should obey
\begin{equation}\label{6.1}
(\nabla_p u)^A_{\cal O} = 0.
\end{equation}
The general class I Szekeres metric \cite{PlKr2006} in the signature $(- + + +)$
is
\begin{equation}\label{6.2}
{\rm d} s^2 = - {\rm d} t^2 + \frac {F^2} {\varepsilon - k(z)} {\rm d} z^2 +
\left(\frac {\Phi} {\cal E}\right)^2 \left({\rm d} x^2 + {\rm d} y^2\right),
\end{equation}
where $(x^1, x^2, x^3) = (z, x, y)$, $\varepsilon = \pm 1, 0$, and
\begin{equation}\label{6.3}
{\cal E} \df \frac S 2 \left[\left(\frac {x - P} S\right)^2 + \left(\frac {y -
Q} S\right)^2 + \varepsilon\right],
\end{equation}
\begin{equation}\label{6.4}
F = \Phi,_z - \Phi {\cal E},_z / {\cal E},
\end{equation}
$P(z)$, $Q(z)$, $S(z)$ and $k(z)$ are arbitrary functions and $\Phi(t, z)$ is
determined by the (generalised Friedmann) equation
\begin{equation}\label{6.5}
{\Phi,_t}^2 = - k(z) + \frac {2 M(z)} {\Phi} - \frac 1 3 \Lambda \Phi^2,
\end{equation}
$M(z)$ being one more arbitrary function and $\Lambda$ being the cosmological
constant. The source in the Einstein equations is dust, with the mass density
\begin{equation}\label{6.6}
\kappa \rho = \frac {2 \left(M,_z - 3 M {\cal E},_z / {\cal E}\right)} {\Phi^2
F}.
\end{equation}

The 2-metric $(\Phi / {\cal E})^2 \left({\rm d} x^2 + {\rm d} y^2\right)$ is
that of a 2-dimensional surface of constant curvature. With $\varepsilon = +1$
it is a sphere, with $\varepsilon = 0$ it is flat (but not necessarily a
Euclidean plane \cite{Kras2008}), with $\varepsilon = -1$ its curvature is
negative. The spheres are in general nonconcentric. The Szekeres metrics
corresponding to the 3 values of $\varepsilon$ are called, respectively,
quasi-spherical, quasi-plane and quasi-hyperbolic (or quasi-pseudospherical).
For more information about the possible Szekeres geometries see Refs.
\cite{Kras2008} -- \cite{KrBo2012}.

The Lema\^{\i}tre -- Tolman limit of (\ref{6.2}) -- (\ref{6.6}) is $\varepsilon
= +1$ and $P, Q, S$ being constant, in the Friedmann limit $k / M^{2/3} =$
constant and the function $t_B(z)$ that appears in the solution of (\ref{6.5})
is also constant.

Now we proceed in the same way as in Sec. \ref{conflat}. In the Szekeres metrics
$u^{\alpha} = {\delta^{\alpha}}_0$, $w^{\alpha} = 0$ and $(\nabla_p u)^{\alpha}
= \Chr {\alpha}\ {0 \rho\ }\ p^{\rho}$, so
\begin{eqnarray}
&& (\nabla_p u)^0 = 0, \qquad (\nabla_p u)^1 = p^1 F,_t / F, \label{6.7} \\
&& (\nabla_p u)^2 = p^2 \Phi,_t / \Phi, \qquad (\nabla_p u)^3 = p^3 \Phi,_t /
\Phi. \label{6.8}
\end{eqnarray}
The equation $q^{\alpha} u_{\alpha} = 0$ at ${\cal O}$ implies $q^0_{\cal O} =
0$, while $q^{\alpha} p_{\alpha} = 0$ means
\begin{equation}\label{6.9}
g_{11} q^1 p^1 + g_{22} q^2 p^2 + g_{33} q^3 p^3 = 0.
\end{equation}
Using (\ref{6.9}) we find
\begin{equation}\label{6.10}
\left(\nabla_p u\right)^q = g_{\alpha \beta} q^{\alpha} \left(\nabla_p
u\right)^{\beta} = g_{11} q^1 p^1 \left(F,_t / F - \Phi,_t / \Phi\right).
\end{equation}
The geodesic equations prohibit $p^1 = 0$ on an open interval of any ray, this
can happen only at isolated points (otherwise the geodesic would be timelike)
\cite{KrBo2011}. The symbol $q^{\alpha}$ represents both tetrad vectors
${e_A}^{\alpha}$ which span a 2-dimensional tangent plane to the spacetime. So,
$q^1 = 0$ can happen for one of them, but not for both at the same point. The
locus of $g_{11} = 0$, if it exists, is a removable coordinate singularity or a
special location called neck \cite{PlKr2006}. So, $\left(\nabla_p u\right)^q =
0$ can hold for all comoving emitter -- comoving observer pairs only when $F,_t
/ F - \Phi,_t / \Phi$, which implies $\Phi = \beta(z) a(t)$, $\beta$ and $a$
being arbitrary functions. Such a form of $\Phi$ ensures that shear of the
cosmic velocity field is zero, but consistency with (\ref{6.5}) requires that in
addition $\beta / M^{1/3}$ and $k / M^{2/3}$ are both constant. In this limit
the Szekeres metrics reduce to the Friedmann models represented in untypical
coordinates. So, the Szekeres metrics obey (\ref{5.5}) only in the Friedmann
limit.

For the class II Szekeres (SII) metrics the result is analogous: (\ref{6.1})
implies zero shear, and in this limit the Friedmann metric results. The formulae
defining the SII metric are long and numerous (they can all be found in
\cite{PlKr2006}), so quoting them to demonstrate this simple result would unduly
expand this paper. Readers are asked to believe or verify.

We proved that the Szekeres metrics cannot obey (\ref{5.5}) except in the
Friedmann limit. But (\ref{5.5}) is a necessary condition for the absence of
drift. So, the final conclusion is that the Szekeres metrics become drift-free
in the KK sense only in the Friedmann limit. This is consistent with Ref.
\cite{KrBo2011}.

\section{Comparison with the results of Ref. [11]}\label{Perlick}

\setcounter{equation}{0}

Spacetimes with zero drift (called there parallax-free world models) were
discussed by Hasse and Perlick (HP) \cite{HaPe1988}. They did not consider the
Einstein equations, they assumed only that a unit timelike vector field $u$
exists that is tangent to the world lines of observers comoving with the cosmic
matter. They defined zero drift as follows: A world model $(M, g, u)$ (where $M$
is the spacetime manifold and $g$ is the metric) is parallax-free if and only
if, for any three observers $a_0$, $a_1$ and $a_2$, the angle under which $a_1$
and $a_2$ are seen by $a_0$ remains constant over time.

Then they proved that this definition is equivalent to 6 other conditions, of
which we quote only the following four:

P2: If $a_0$ sees $a_1$ and $a_2$ in the same spatial direction at one instant,
then $a_1$ and $a_2$ will stay in the same spatial direction at all instants.

P3: The vector field $u$ is proportional to a conformal Killing field.

P4: There is some scalar function $f$ on $M$ such that
\begin{equation}\label{7.1}
\lie u {g_{\alpha \beta}} = - 2 u^{\rho} f,_{\rho} g_{\alpha \beta} + u_{\alpha}
f,_{\beta} + u_{\beta} f,_{\alpha}.
\end{equation}

P5: $u^{\alpha}$ is shearfree and the one-form $\omega$ defined by
\begin{equation}\label{7.2}
c^2 \omega \df \dot{u}_{\alpha} {\rm d} x^{\alpha} - \tfrac 1 3\ \theta
u_{\alpha} {\rm d} x^{\alpha},
\end{equation}
where $\dot{u}^{\alpha} \df {u^{\alpha}};_{\rho} u^{\rho}$ and $\theta \df
{u^{\rho}};_{\rho}$, satisfies ${\rm d} \omega = 0$.

We now verify P2 -- P5 in the Stephani metric (\ref{5.6}) -- (\ref{5.7}).

A conformal Killing vector field $k^{\alpha}$ obeys, by definition
\begin{equation}\label{7.3}
k^{\rho} g_{\alpha \beta, \rho} + {k^{\rho}},_{\alpha} g_{\rho \beta} +
{k^{\rho}},_{\beta} g_{\alpha \rho} = \mu g_{\alpha \beta},
\end{equation}
where $\mu$ is a scalar function.\footnote{All the partial derivatives in
(\ref{7.3}) can be replaced by covariant derivatives because the terms
containing the Christoffel symbols cancel out. Then (\ref{7.3}) assumes the more
familiar form $k_{\alpha; \beta} + k_{\beta; \alpha} = \mu g_{\alpha \beta}$.}
The comoving observer velocity field in the metric (\ref{5.6}) -- (\ref{5.7}) is
given by (\ref{5.10}), so
\begin{equation}\label{7.4}
k^{\alpha} = \frac V {\phi F V,_t}\ {\delta^{\alpha}}_0
\end{equation}
should obey (\ref{7.3}) in those subcases of the Stephani metric that are
parallax-free by criterion P3; $\phi$ is a function to be determined. The labels
of the coordinates will be $(x^0, x^1, x^2, x^3) = (t, x, y, z)$.

With (\ref{5.6}) and (\ref{7.4}), the components $(I, I)$, $I = 1, 2, 3$, of
(\ref{7.3}) imply
\begin{equation}\label{7.5}
\mu = - \frac 2 {\phi F},
\end{equation}
and the components (0, I) of (\ref{7.3}) imply
\begin{equation}\label{7.6}
k^0 = \frac V {\phi F V,_t} = G(t),
\end{equation}
where $G(t)$ is an arbitrary function. The component (0, 0) of (\ref{7.3}) now
is
\begin{equation}\label{7.7}
F \frac {V,_t} V\ \left(\frac {V,_t} V - \frac {\phi,_t} {\phi}\right) = 0.
\end{equation}
The metric (\ref{5.6}) does not allow the limit $F V,_t = 0$, so (\ref{7.7})
implies
\begin{equation}\label{7.8}
\phi = V / \gamma(x, y, z),
\end{equation}
where $\gamma$ is a function of $x, y$ and $z$ to be determined. Now (\ref{7.6})
becomes
\begin{equation}\label{7.9}
V,_t = \frac 1 {F(t) G(t)}\ \gamma.
\end{equation}
This implies
\begin{equation}\label{7.10}
\frac {V,_{tt}} {V,_t} = - \frac {(GF),_t} {GF} \df \alpha(t) \Longrightarrow
V,_{tt} = \alpha(t) V,_t.
\end{equation}
With $V$ given by (\ref{5.7}) the above becomes a quadratic equation in $x, y,
z$ with coefficients depending on $t$. We discard the trivial case $k = 0$,
which is the spatially flat RW metric. The coefficients of the various powers of
$x, y$ and $z$ in (\ref{7.10}) must cancel out separately. The coefficients of
$x^2$, $y^2$ and $z^2$ cancel out when $(k / R),_t = 0$ (we consider this case
further on) or
\begin{equation}\label{7.11}
\alpha = (k / R),_{tt} / (k / R),_t.
\end{equation}
For now, we follow (\ref{7.11}) and assume, for the beginning, $x_{0, t} \neq 0
\neq y_{0, t}$, $z_{0, t} \neq 0$. Then, the remaining equations in (\ref{7.10})
imply
\begin{equation}\label{7.12}
\frac {x_{0, tt}} {x_{0, t}} = \frac {y_{0, tt}} {y_{0, t}} = \frac {z_{0, tt}}
{z_{0, t}} = \frac {(k / R)_{tt}} {(k / R)_t} - 2\ \frac {(k / R)_t} {k / R}\ .
\end{equation}
Now the terms linear in $x, y, z$ cancel out in (\ref{7.10}). The solutions of
(\ref{7.12}) are
\begin{equation}\label{7.13}
x_0 = C_1 R / k + D_1, \quad y_0 = C_2 R / k + D_2, \quad z_0 = C_3 R / k + D_3.
\quad
\end{equation}

The cases $x_{0,t} = 0$, $y_{0,t} = 0$ and $z_{0,t} = 0$ are contained in
(\ref{7.13}) as the subcases $C_1 = 0$, $C_2 = 0$ and $C_3 = 0$, respectively.
When $C_1 = C_2 = C_3 = 0$, the metric (\ref{5.6}) becomes spherically
symmetric, but not identical with Robertson -- Walker as long as $k,_t \neq 0$.

The last equation of (\ref{7.10}) not yet taken into account defines $R$; it is
\begin{equation}\label{7.14}
\left(\frac 1 R\right),_{tt} + \frac k {2R}\ \left({x_{0,t}}^2 + {y_{0,t}}^2 +
{z_{0,t}}^2\right) = \frac {(k / R)_{tt}} {(k / R)_t}\ \left(\frac 1
R\right),_t,
\end{equation}
and its solution is
\begin{equation}\label{7.15}
\frac 1 R = - \frac 1 4\ \frac {{C_1}^2 + {C_2}^2 + {C_3}^2} {k / R}\ + E_1
\frac k R + E_2,
\end{equation}
where $E_1$ and $E_2$ are arbitrary constants and $k(t)$ remains arbitrary (but
$\neq 0$). With (\ref{7.13}) and (\ref{7.15}), Eq. (\ref{7.9}) is fulfilled and:
\begin{equation}\label{7.16}
\frac 1 {FG} = \left(\frac k R\right),_t, \qquad \gamma = E_1 + \frac  1 4\
\left[\left(x - D_1\right)^2 + \left(y - D_2\right)^2 + \left(z -
D_3\right)^2\right].
\end{equation}

The $k \neq 0$ Robertson -- Walker models are contained in (\ref{7.15}) as the
subcase $C_1 = C_2 = C_3 = 0$, $E_1 = 1 / k$, $E_2 = 0$ -- then (\ref{7.15})
becomes an identity and does not determine $R(t)$, which remains arbitrary.

With (\ref{7.13}) and (\ref{7.15}) fulfilled, the Stephani metric becomes
axially symmetric and coincides with the subcase found in Ref. \cite{Kras2011}
to have all light paths repeatable (see Appendix A to \cite{Kras2011}, case
1.2.1.2). This shows that the criteria of zero drift of \cite{Kras2011} and
\cite{HaPe1988} coincide for the Stephani metric (\ref{5.6}) -- (\ref{5.7}).
(See Sec. \ref{compa} -- the two criteria coincide in general.)

In the case $(k / R),_t = 0$ that was left aside at (\ref{7.11}), Eq.
(\ref{7.10}) implies
\begin{equation}\label{7.17}
x_{0, tt} = \alpha x_{0, t}, \qquad y_{0, tt} = \alpha y_{0, t}, \qquad z_{0,
tt} = \alpha z_{0, t},
\end{equation}
\begin{equation}\label{7.18}
\left(\frac 1 R\right),_{tt} + \frac k {2R}\ \left({x_{0,t}}^2 + {y_{0,t}}^2 +
{z_{0,t}}^2\right) = \alpha(t) \left(\frac 1 R\right),_t,
\end{equation}
This subcase is also axially symmetric, but was not displayed in
\cite{Kras2011}.

Now we verify that HP's condition P4, i.e., Eq. (\ref{7.1}), is equivalent to
P3. Written out explicitly, (\ref{7.1}) says
\begin{equation}\label{7.19}
u^{\rho} g_{\alpha \beta, \rho} + {u^{\rho}},_{\alpha} g_{\rho \beta} +
{u^{\rho}},_{\beta} g_{\alpha \rho} =  - 2 u^{\rho} f,_{\rho} g_{\alpha \beta} +
u_{\alpha} f,_{\beta} + u_{\beta} f,_{\alpha}.
\end{equation}
We substitute
\begin{equation}\label{7.20}
f = \ln \phi,
\end{equation}
and then (\ref{7.19}) can be rewritten as
\begin{equation}\label{7.21}
\frac {u^{\rho}} {\phi}\ g_{\alpha \beta, \rho} + \left(\frac {u^{\rho}}
{\phi}\right),_{\alpha} g_{\rho \beta} + \left(\frac {u^{\rho}}
{\phi}\right),_{\beta} g_{\alpha \rho} = - 2 u^{\rho} f,_{\rho} g_{\alpha
\beta},
\end{equation}
which shows that $u^{\alpha} / \phi$ is a conformal Killing field, just as in
HP's condition P3. If we repeat for (\ref{7.21}) the reasoning previously
applied to (\ref{7.3}), then we will find that $2 u^{\rho} \phi,_{\rho} / \phi^2
= 2 u^0 V,_t / (\phi V) = 2 / (\phi F)$. Thus indeed P4 applied to the expanding
Stephani metric is equivalent to P3.

HP's condition P5, i.e., ${\rm d} \omega = 0$ for $\omega$ given by (\ref{7.2}),
is satisfied nearly trivially for the subcase of the Stephani metric given by
(\ref{7.13}) -- (\ref{7.15}). To see this, one must use $\theta = - 3 / F$,
$\dot{u}_0 = 0$, $\dot{u}_I = - V,_{tI} / V,_t + V,_I / V$, $I = 1, 2, 3$, and
(\ref{7.10}).

\section{Comparison with the results of Refs.
[4, 7]}\label{compa}

\setcounter{equation}{0}

In the KB paper \cite{KrBo2011} the zero-drift condition (called there the
condition for repeatable light paths) was that all light rays sent from any
fixed comoving emitter to any fixed comoving observer intersect the same set of
intermediate world lines of cosmic matter. This is clearly equivalent to HP's
condition P2 (see Sec. \ref{Perlick}). Consequently, it is not surprising that
the HP and KB criteria of zero drift applied to the Stephani metric selected the
same subcase (\ref{7.13}) -- (\ref{7.15}), see Ref. \cite{Kras2011}.

In the language of Ref. \cite{KoKo2018} the KB = HP definitions mean that the
world lines of cosmic matter lie in the surfaces $P_2$ tangent to the
observation time vector field $X^{\mu}$ and to the geodesic field $p^{\mu}$, and
actually are everywhere tangent to the field $X^{\mu}$. The first of (\ref{3.7})
must hold at the initial point of every past-directed ray, the second of
(\ref{3.7}) must hold all along that ray.

The condition (\ref{4.5}) (which follows from (\ref{5.5}) via (\ref{4.1})),
together with the requirement that the metric obeys the Einstein equations with
a perfect fluid source, leads to the conclusion that the spacetime must be
conformally flat. The conformally flat perfect fluid metrics are explicitly
known, they are the Stephani metrics \cite{Step1967, Step2003}, see our Sec.
\ref{conflat}. So, it was natural to verify whether the reverse implication also
occurs. Indeed, it was proved in Sec. \ref{conflat} that the Stephani metric in
its full generality obeys (\ref{5.5}), and then, in Appendix B, that it obeys
(\ref{4.4}). But, as noted below (\ref{4.5}), Eq. (\ref{4.5}) is only a
necessary condition for the absence of the drift in the KK sense -- it does not
represent the whole information contained in (\ref{4.2}). The perfect fluid
metric that is drift-free in the KK sense is the subcase of the Stephani
solution given by (\ref{7.13}) -- (\ref{7.15}).

Let us recall what was said at the beginning of Sec. \ref{implizerdri}: When
$\delta_{\cal O} r^A = 0$, the whole 4-dimensional $\delta_{\cal O} r^{\mu} =
0$. Then, angles between the directions to any pair of light sources will remain
constant in observer's time. This means that the spacetimes that are drift-free
in the KK sense are necessarily drift-free also in the HP sense.

\begin{figure}[h]
\begin{center}
\hspace{-5cm} \includegraphics[scale = 0.9]{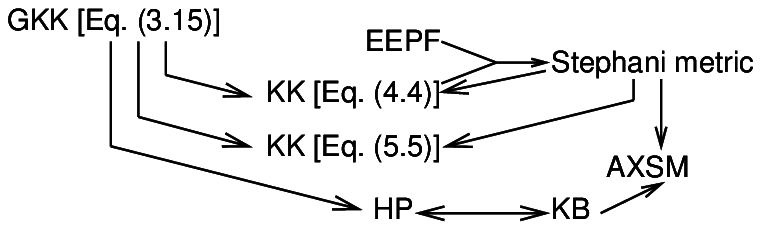}
 \caption{Relations between the results reported in this paper. Arrows show
implications. See the text for explanations of the abbreviations.}
 \label{diagram}
\end{center}
\end{figure}

The relations between the KK, HP, KB approaches and the Stephani metric are
briefly summarised in Fig. \ref{diagram}. Here are the explanations of the
abbreviations used in Fig. \ref{diagram}:
\begin{itemize}
\item EEPF = Einstein equations with perfect fluid source.
\item Stephani metric = the metric given by (\ref{5.6}) -- (\ref{5.7}).
\item {\underline {The criteria for absence of drift}}:
\item GKK = the general Korzy\'nski -- Kopi\'nski (2018) criterion
(\ref{3.15}) with $\delta_{\cal O} r^A = 0$.
\item KK [Eq. (\ref{4.4})] = the conclusion (\ref{4.4}) from the KK criterion.
\item KK [Eq. (\ref{5.5})] = the conclusion (\ref{5.5}) from the KK criterion.
\item HP = the set of Hasse -- Perlick (1988) equivalent criteria.
\item KB = the Krasi\'nski -- Bolejko (2011) criterion.
\item AXSM = the axially symmetric subcase of the Stephani metric given by
(\ref{5.6}) -- (\ref{5.7}) with (\ref{7.13}) -- (\ref{7.15}).
\end{itemize}

It is simple to verify that HP's condition P2 of Sec. \ref{Perlick}, i.e., the
existence of a conformal Killing field collinear with the velocity field
$u^{\alpha} = {\delta^{\alpha}}_0$, imposed on the class I Szekeres metric,
implies zero shear, i.e., the Friedmann limit. Thus, for the Szekeres metrics of
class I all three approaches give the same result: the position drift vanishes
for all comoving observer -- comoving emitter pairs only in the Friedmann limit.

\section{Summary and conclusions}\label{summa}

\setcounter{equation}{0}

Light rays proceeding through the Universe cross evolving condensations and
voids, where the cosmic matter may move with shear and rotation. All those
encounters cause deflections of the rays, and in general each angle of
deflection changes with time. As a consequence, a typical observer should see
the direction to each given light source change (in practice, very slowly) with
time. This effect is referred to as position drift \cite{KoKo2018}. A few teams
of authors noted the necessity of taking this drift into account, from the point
of view of both theory \cite{KrBo2011,Kras2011,Kras2012}, \cite{HaPe1988} --
\cite{KMSe2021} and observations \cite{QABC2012}. The theoretical approaches
were by two methods: checking whether light rays reaching the observer proceed
from a given light source through always the same intermediate world lines of
cosmic medium (the HP \cite{HaPe1988} and KB \cite{KrBo2011} approaches) and
calculating the change of direction toward a given source with respect to a
reference plane (the Fermi -- Walker derivative of the direction vector along
the observer world line, the KK \cite{KoKo2018} approach). The aim of the
present paper was to compare the methods and results of these three approaches.
It turned out that the HP and KB criteria of zero drift are equivalent, and are
a necessary condition for the KK criterion to apply. The expanding Stephani
metric is drift-free by the HP = KB criterion when its metric functions obey the
additional conditions (\ref{7.13}) -- (\ref{7.15}). With these conditions
fulfilled, it becomes axially symmetric and contains the general Robertson --
Walker metric (\ref{4.12}) as a still more special case.

In detail, these are the results of the present paper.

Sections \ref{intro} -- \ref{posdrift} contain the motivation (Sec.
\ref{intro}), the formulae for basic quantities expressed in the semi-null
tetrad defined by the observer velocity and a light ray (Sec. \ref{SNT}), and a
summary of the KK approach (Sec. \ref{posdrift}).

In Sec. \ref{implizerdri} it is shown that if the spacetime metric obeys the
Einstein equations with a perfect fluid source (this assumption underlies all
relativistic cosmological models), then the necessary condition for zero
position drift by the KK definition for all comoving observers is conformal
flatness of the metric. Details of the calculation are given in Appendix A.

The most general conformally flat perfect fluid solutions of the Einstein
equations are explicitly known, they are the Stephani metrics
\cite{Step1967,Step2003}. In Sec. \ref{conflat} it is shown that the general
Stephani metric obeys Eq. (\ref{5.5}), which is a necessary condition for zero
drift by the KK definition, and in Appendix B it is shown that the general
Stephani metric also obeys another necessary condition for KK zero drift, namely
Eq. (\ref{4.4}).

In Sec. \ref{Szek} it is shown that the Szekeres metrics
\cite{Szek1975,Szek1975b,PlKr2006} become drift-free in the KK sense only in the
Friedmann limit.

In Sec. \ref{Perlick} it is shown that the KB condition of zero drift
\cite{KrBo2011} coincides with the HP condition \cite{HaPe1988}. It is also
shown that the HP condition imposed on the Stephani metric
\cite{Step1967,Step2003} leads to the same subcase as the one identified as
drift-free in Ref. \cite{Kras2011}.

Finally, in Sec. \ref{compa} the relations between the HP, KB and KK approaches
are compared, discussed and explained, and the relations between the various
results of this paper are shown in a graphic diagram.

\bigskip

{\bf Acknowledgements.} For some calculations, the computer algebra system
Ortocartan \cite{Kras2001,KrPe2000} was used. I am grateful to the referee for
pointing out an incompleteness of the first version of this paper.

\appendix

\section{Full implications of Eq. (4.11) }

\setcounter{equation}{0}

All indices appearing in this appendix will be tetrad indices, so for
transparency the hats above them are omitted.

Equation (\ref{4.11}) was derived in the tetrad adapted to that null vector
field $p^{\alpha}$ which is tangent to the family of null geodesics connecting a
fixed light emitter and a fixed observer, both comoving with the cosmic fluid.
In that tetrad, the frame components of $p^{\alpha}$ are $p^i = \delta^i_3$, as
in (\ref{2.17}).

Now consider another null vector $q^i$ tangent to another ray reaching the same
observer ${\cal O}$. With the tetrad metric given by (\ref{2.4}) -- (\ref{2.6}),
the condition for $q^i$ to be null is
\begin{equation}\label{a.1}
0 = \eta_{i j} q^i q^j = \left(q^0\right)^2 + 2 u_{\rho} p^{\rho} q^0 q^3 -
\left(q^1\right)^2 - \left(q^2\right)^2,
\end{equation}
and we wish to find all implications of
\begin{equation}\label{a.2}
{C^A}_{i j 0} q^i q^j = 0
\end{equation}
for all $q^i$ obeying (\ref{a.1}).

By virtue of (\ref{4.11}) we have
\begin{equation}\label{a.3}
C_{0 3 1 3} = C_{0 3 2 3} = 0.
\end{equation}

Equation (\ref{4.11}) is (\ref{a.2}) applied to $q^i = p^i = \delta^i_3$, which
is the only nontrivial solution of (\ref{a.1}) with $q^0 = 0$. When $q^0 \neq
0$, at least one other component of $q^i$ must be nonzero. Let us begin with the
case
\begin{equation}\label{a.4}
q^1 = q^2 = 0, \qquad q^0 = - 2 u_{\rho} p^{\rho} q^3.
\end{equation}
In consequence of (\ref{4.11}), eq. (\ref{a.2}) becomes in this case ${C^A}_{0 3
0} q^0 q^3 = 0$, so
\begin{equation}\label{a.5}
C_{0 1 0 3} = C_{0 2 0 3} = 0.
\end{equation}
Now let us take
\begin{equation}\label{a.6}
q^1 = q^3 = 0, \qquad q^0 = \pm q^2 \neq 0.
\end{equation}
Then (\ref{a.2}) becomes
\begin{equation}\label{a.7}
\left(q^2\right)^2 \left(\pm {C^A}_{0 2 0} + {C^A}_{2 2 0}\right) = 0,
\end{equation}
and this must hold for both signs. Consequently
\begin{equation}\label{a.8}
C_{0 1 0 2} = C_{0 2 0 2} = C_{0 2 1 2} = 0.
\end{equation}
Now we take
\begin{equation}\label{a.9}
q^2 = q^3 = 0, \qquad q^0 = \pm q^1 \neq 0
\end{equation}
and use this in (\ref{a.2}):
\begin{equation}\label{a.10}
\left(q^1\right)^2 \left(\pm {C^A}_{0 1 0} + {C^A}_{1 1 0}\right) = 0.
\end{equation}
{}From here, the following new equations result:
\begin{equation}\label{a.11}
C_{0 1 0 1} = C_{0 1 1 2} = 0.
\end{equation}
We take now $q^2 = 0$, all other $q^i \neq 0$. Using the information about $C_{i
j k l}$ gained up to now, the equation ${C^A}_{i j 0} q^i q^j = 0$ reduces to
\begin{equation}\label{a.12}
q^1 q^3 \left({C^A}_{3 1 0} + {C^A}_{1 3 0}\right) = 0,
\end{equation}
which implies $C_{A 3 1 0} + C_{A 1 3 0} = 0$, and then
\begin{eqnarray}
C_{0 1 1 3} &=& 0, \label{a.13} \\
- C_{0 1 2 3} + C_{0 3 1 2} &=& 0.  \label{a.14}
\end{eqnarray}
Now let $q^1 = 0$, all other $q^i \neq 0$. Using what we know about $C_{i j k
l}$ we get
\begin{equation}\label{a.15}
q^2 q^3 \left({C^A}_{3 2 0} + {C^A}_{2 3 0}\right) = 0,
\end{equation}
and lowering the index $A = 1, 2$ we get from here:
\begin{eqnarray}
C_{0 2 2 3} &=& 0, \label{a.16} \\
C_{0 2 1 3} + C_{0 3 1 2} &=& 0.  \label{a.17}
\end{eqnarray}
Now (\ref{a.14}) and (\ref{a.17}) imply
\begin{equation}\label{a.18}
C_{0 1 2 3} = - C_{0 2 1 3}.
\end{equation}
Let us now use the identity $C_{i j k l} + C_{i k l j} + C_{i l j k} = 0$:
\begin{equation}\label{a.19}
C_{0 1 2 3} - C_{0 2 1 3} + C_{0 3 1 2} = 0.
\end{equation}
Together with (\ref{a.14}) and (\ref{a.17}) this implies
\begin{equation}\label{a.20}
C_{0 1 2 3} = 0,
\end{equation}
and then (\ref{a.18}) with (\ref{a.17}) imply
\begin{equation}\label{a.21}
C_{0 2 1 3} = C_{0 3 1 2} = 0.
\end{equation}

In further calculations, the following formulae will be needed (in the last one
we made use of Eqs. (\ref{a.5}) -- (\ref{a.3}) which showed that $C_{0 A k l} =
0$ for both $A$ and all $k, l$):
\begin{eqnarray}
{C^{0 A}}_{k l} &=& \frac 1 {u_{\rho} p^{\rho}}\ C_{A 3 k l}, \qquad {C^{0
3}}_{k l} = - \frac 1 {\left(u_{\rho} p^{\rho}\right)^2}\ C_{0 3 k l}, \nonumber
\\
{C^{1 2}}_{k l} &=& C_{1 2 k l}, \qquad {C^{A 3}}_{k l} = \frac 1
{\left(u_{\rho} p^{\rho}\right)^2}\ C_{A 3 k l}.  \label{a.22}
\end{eqnarray}
Now we use ${C^{i j}}_{k j} = 0$. From ${C^{0 j}}_{0 j} = 0$, using (\ref{a.16})
and (\ref{a.13}) we get
\begin{equation}\label{a.23}
C_{0 3 0 3} = 0.
\end{equation}
{}From ${C^{0 j}}_{1 j} = 0$, using (\ref{a.3}) we get
\begin{equation}\label{a.24}
C_{1 2 2 3} = 0.
\end{equation}
{}From ${C^{0 j}}_{2 j} = 0$, using (\ref{a.3}) we get
\begin{equation}\label{a.25}
C_{1 2 1 3} = 0.
\end{equation}
{}From ${C^{0 j}}_{3 j} = 0$ we get
\begin{equation}\label{a.26}
C_{1 3 1 3} + C_{2 3 2 3} = 0.
\end{equation}
{}From ${C^{1 j}}_{1 j} = 0$ we get
\begin{equation}\label{a.27}
C_{1 2 1 2} + \frac 1 {\left(u_{\rho} p^{\rho}\right)^2}\ C_{1 3 1 3} = 0.
\end{equation}
{}From ${C^{1 j}}_{2 j} = 0$, using (\ref{a.21}) we get
\begin{equation}\label{a.28}
C_{1 3 2 3} = 0.
\end{equation}
{}From ${C^{2 j}}_{2 j} = 0$ using (\ref{a.16}) we get
\begin{equation}\label{a.29}
C_{1 2 1 2} + \frac 1 {\left(u_{\rho} p^{\rho}\right)^2}\ C_{2 3 2 3} = 0.
\end{equation}
{}From (\ref{a.27}) and (\ref{a.29}) we see that
\begin{equation}\label{a.30}
C_{1 3 1 3} = C_{2 3 2 3}.
\end{equation}
{}From ${C^{3 j}}_{3 j} = 0$ we get, using (\ref{a.23})
\begin{equation}\label{a.31}
C_{1 3 1 3} + C_{2 3 2 3} = 0.
\end{equation}
Together with (\ref{a.30}) this means
\begin{equation}\label{a.32}
C_{1 3 1 3} = C_{2 3 2 3} = 0,
\end{equation}
and then (\ref{a.29}) implies
\begin{equation}\label{a.33}
C_{1 2 1 2} = 0.
\end{equation}
At this point, we have shown that all $C_{i j k l} = 0$. The remaining trace
equations bring no new information. $\square$

\section{The expanding Stephani metric obeys Eq. (4.4) with $f = 0$. }

\setcounter{equation}{0}

With (\ref{5.6}) -- (\ref{5.7}) every null vector must obey
\begin{equation}\label{b.1}
D^2 \left(p^0\right)^2 = \frac 1 {V^2}\ \left[\left(p^1\right)^2 +
\left(p^2\right)^2 + \left(p^3\right)^2\right], \qquad D \df F V,_t / V.
\end{equation}
The only nonzero components of the Riemann tensor are those given below, plus
those related to them by the simple indicial symmetries:
\begin{eqnarray}
R_{0 1 0 1} &=& R_{0 2 0 2} = R_{0 3 0 3} = \frac 1 {V^2} \left(C^2 D^2 - F C
C,_t D\right) \df A \label{b.2} \\
R_{1 2 1 2} &=& R_{1 3 1 3} = R_{2 3 2 3} = - C^2 / V^4. \label{b.3}
\end{eqnarray}
We recall that the cosmic velocity field $u^{\alpha}$ (given by (\ref{5.10}))
has only the $u^0$ component, and so
\begin{equation}\label{b.4}
X^{\alpha} = \frac {u^{\alpha}} {1 + z} = \frac {B u^{\alpha}} {p_{\sigma}
u^{\sigma}} = \frac {B u^0} {p_0 u^0}\ \delta^{\alpha}_0, \qquad B \df
\left(p_{\sigma} u^{\sigma}\right)_{\cal O}
\end{equation}
(the $B$ is constant along each ray, but different on different rays), and
further
\begin{equation}\label{b.5}
X^{\alpha} = \frac B {p_0}\ \delta^{\alpha}_0 = \frac B {D^2 p^0}\
\delta^{\alpha}_0.
\end{equation}
Writing (\ref{4.4}) with the index $\mu = 0$ we obtain
\begin{equation}\label{b.6}
\left(f g + \nabla_p g\right) p^0 = \frac 1 {D^2}\ R_{0 \alpha \beta 0}\
p^{\alpha} p^{\beta}\ \frac B {D^2 p^0}.
\end{equation}
In view of (\ref{b.2}) this is equivalent to
\begin{equation}\label{b.7}
\frac 1 B\ \left(f g + \nabla_p g\right) \left(p^0\right)^2 = - \frac A {D^4}\
\left[\left(p^1\right)^2 + \left(p^2\right)^2 + \left(p^3\right)^2\right],
\end{equation}
and then (\ref{b.1}) gives
\begin{equation}\label{b.8}
\left(f g + \nabla_p g\right) / B = - A V^2 / D^2 = C C,_t V / V,_t - C^2.
\end{equation}
This defines $\left(f g + \nabla_p g\right) / B$ in terms of the metric
functions. The remaining 3 equations in (\ref{4.4}) are either fulfilled
identically or just duplicate (\ref{b.8}). As an example, let us take
(\ref{4.4}) with $\mu = 1$:
\begin{equation}\label{b.9}
\left(f g + \nabla_p g\right) p^1 = {R^1}_{\alpha \beta 0} p^{\alpha} p^{\beta}\
B / \left(D^2 p^0\right).
\end{equation}
Using (\ref{b.2}) -- (\ref{b.3}), the only nonzero terms in (\ref{b.9}) are
those with $\alpha = 0, \beta = 1$, so
\begin{equation}\label{b.10}
\frac 1 B\ \left(f g + \nabla_p g\right) p^1 = - \left(A V^2 / D^2\right) p^1.
\end{equation}
If $p^1 = 0$, then (\ref{b.10}) is an identity; if $p^1 \neq 0$, then
(\ref{b.10}) duplicates (\ref{b.8}).

For the next equations we will need the expressions for the following
Christoffel symbols:
\begin{eqnarray}
\Chr{I\ }\ {0 0\ } &=& V^2 D D,_I, \qquad \Chr{I\ }\ {0 J\ } = - \frac D F\
{\delta^I}_J, \label{b.11} \\
\Chr {0\ }\ {0 \mu\ } &=& \frac {D,_{\mu}} D, \qquad \Chr {0\ }\ {I J\ } = -
\frac 1 {F V^2 D}\ \delta_{I J}.\label{b.12}
\end{eqnarray}

Now let us consider Eq. (\ref{4.1}) for the expanding Stephani metric. Its
spatial components, $\mu = I = 1, 2, 3$, in view of (\ref{b.5}), are equivalent
to
\begin{equation}\label{b.13}
V^2 D,_I / D = g_F p^I, \qquad g_F \df g / B + 1 / \left(F D p^0\right).
\end{equation}
Using (\ref{b.1}) this implies
\begin{equation}\label{b.14}
g_F^2 \left[\left(p^1\right)^2 + \left(p^2\right)^2 + \left(p^3\right)^2\right]
= \frac {V^4} {D^2} \left({D,_x}^2 + {D,_y}^2 + {D,_z}^2\right) = g_F^2 V^2 D^2
\left(p^0\right)^2.
\end{equation}
Using (\ref{b.11}) -- (\ref{b.13}), (\ref{b.1}), (\ref{b.5}) and $p^{\rho}
{p^0},_{\rho} = - \Chr {0\ }\ {\sigma \rho} p^{\sigma} p^{\rho}$ we get from the
$\mu = 0$ component of (\ref{4.1})
\begin{equation}\label{b.15}
\frac {p^I D,_I} {D^2 p^0} - \frac 1 F = \frac f {D p^0} + \frac g B\ D p^0.
\end{equation}
Using (\ref{b.13}) and (\ref{b.14}) and again (\ref{b.1}) this implies
\begin{equation}\label{b.16}
f = 0,
\end{equation}
so (\ref{4.1}) and (\ref{4.4}) take the simpler form
\begin{eqnarray}
\nabla_p X^{\mu} &=& g p^{\mu}  \label{b.17} \\
\left(\nabla_p g\right) p^{\mu} &=& {R^{\mu}}_{\alpha \beta \nu} p^{\alpha}
p^{\beta} X^{\nu}.\label{b.18}
\end{eqnarray}


\begin{thebibliography}{99}
\bibitem{Szek1975} P. Szekeres, A class of inhomogeneous cosmological models.
{\it Commun. Math. Phys.} {\bf 41}, 55 (1975).

\bibitem{Szek1975b} P. Szekeres, Quasispherical gravitational collapse. {\it
Phys. Rev.} {\bf D12}, 2941 (1975).

\bibitem{PlKr2006} J. Pleba\'nski and A. Krasi\'nski, {\it An Introduction to
General Relativity and Cosmology}. Cambridge University Press 2006.

\bibitem{KrBo2011} A. Krasi\'nski and K. Bolejko, Redshift propagation
equations in the $\beta' \neq 0$ Szekeres models. {\it Phys. Rev.} {\bf D83},
083503 (2011).

\bibitem {Barn1973} A. Barnes, On shearfree normal flows of a perfect
fluid, {\it Gen. Relativ. Gravit.} {\bf 4}, 105 (1973).

\bibitem{Step1967} H. Stephani, \"{U}ber L\"{o}sungen der Eisteinschen
Feldgleichungen, die sich in einen f\"{u}nfdimensionalen flachen Raum einbetten
lassen [On solutions of Einstein's equations that can be embedded in a
five-dimensional flat space], {\it Commun. Math. Phys}. {\bf 4}, 137 (1967).

\bibitem{Kras2011} A. Krasi\'nski, Repeatable light paths in the shearfree
normal cosmological models. {\it Phys. Rev.} {\bf D84}, 023510 (2011).

\bibitem{Kras2012} A. Krasi\'nski, Repeatable light paths in the conformally
flat cosmological models. {\it Phys. Rev.} {\bf D86}, 064001 (2012).

\bibitem{Lema1933} G. Lema\^{\i}tre, L'Univers en expansion [The expanding
Universe]. {\it Ann. Soc. Sci. Bruxelles} {\bf A53}, 51 (1933); English
translation: {\it Gen. Relativ. Gravit.} {\bf 29}, 641 (1997); with an editorial
note by A. Krasi\'nski: {\it Gen. Relativ. Gravit.} {\bf 29}, 637 (1997).

\bibitem{Tolm1934} R. C. Tolman, Effect of inhomogeneity on cosmological
models. {\it Proc. Nat. Acad. Sci. USA} {\bf 20}, 169 (1934); reprinted: {\it
Gen. Relativ. Gravit.} {\bf 29}, 935 (1997); with an editorial note by A.
Krasi\'nski, in: {\it Gen. Relativ. Gravit.} {\bf 29}, 931 (1997).

\bibitem{HaPe1988} W. Hasse and V. Perlick, Geometrical and kinematical
characterization of parallax-free world models. {\it J. Math. Phys.} {\bf 29},
2064 (1988).

\bibitem{KoKo2018} M. Korzy\'nski and J. Kopi\'nski, Optical drift effects in
general relativity. {\it Journal of Cosmology and Astroparticle Physics} {\bf
03}, 012 (2018).

\bibitem{GKSe2019} M. Grasso, M. Korzy\'nski and J. Serbenta, Geometric optics
in general relativity using bilocal operators. {\it Phys. Rev.} {\bf D99},
064038 (2019).

\bibitem{KMSe2021} M. Korzy\'nski, J. Mi\'skiewicz and J. Serbenta, Weighing the
spacetime along the line of sight using times of arrival of electromagnetic
signals. {\it Phys. Rev.} {\bf D104}, 024026 (2021).

\bibitem{Step2003} H. Stephani, D. Kramer, M. MacCallum, C. Hoenselaers and E.
Herlt, {\it Exact Solutions of Einstein's Field Equations}. 2nd Edition.
Cambridge University Press (2003).

\bibitem{Rind2011} W. Rindler, {\it Relativity, special, general and
cosmological}, second edition. Oxford University Press, Oxford 2011.

\bibitem{Kras1981} A. Krasi\'nski, Spacetimes with spherically symmetric
hypersurfaces. {\it Gen. Relativ. Gravit.} {\bf 13}, 1021 (1981).

\bibitem{Kras2008} A. Krasi\'nski, Geometry and topology of the quasi-plane
Szekeres model. {\it Phys. Rev.} {\bf D78}, 064038 (2008).

\bibitem{HeKr2002} C. Hellaby and A. Krasi\'nski, You cannot get through
Szekeres wormholes: Regularity, topology and causality in quasi-spherical
Szekeres models. {\it Phys. Rev.} {\bf D66}, 084011 (2002).

\bibitem{HeKr2008} C. Hellaby and A. Krasi\'nski, Physical and geometrical
interpretation of the $\epsilon \leq 0$ Szekeres Models. {\it Phys. Rev.} {\bf
D77}, 023529 (2008).

\bibitem{KrBo2012} A. Krasi\'nski and K. Bolejko, Apparent horizons in the
quasi-spherical Szekeres models. {\it Phys. Rev.} {\bf D85}, 124016 (2012).

\bibitem{QABC2012} C. Quercellini, L. Amendola, A. Balbi, P. Cabella, M.
Quartin, Real-time cosmology. {\it Phys. Reports}. {\bf 521}, 95 -- 134 (2012).

\bibitem{Kras2001} A. Krasi\'nski, The newest release of the Ortocartan set of
programs for algebraic calculations in relativity. {\it Gen. Relativ. Gravit.}
{\bf 33}, 145 (2001).

\bibitem{KrPe2000} A. Krasi\'nski, M. Perkowski, {\it The system ORTOCARTAN --
user's manual}. Fifth edition, Warsaw 2000.
\end{thebibliography}
\end{document}